\documentstyle[times,balanced,epsf,prl,aps,amsfonts]{revtex}

\newcommand{\R}{\Bbb R}
\newcommand{\Z}{\Bbb Z}

\begin{document}

\draft

\title{Universality in Three-Frequency Resonances}

\author{
Julyan H. E. Cartwright$^{1,}$\cite{jemail}, 
Diego L. Gonz\'alez$^{2,}$\cite{demail}, and 
Oreste Piro$^{3,}$\cite{oemail}
} 

\address{
$^1$Instituto Andaluz de Ciencias de la Tierra, IACT (CSIC-UGR),
E-18071 Granada, Spain. \\
$^2$Istituto Lamel, CNR, I-40129 Bologna, Italy \\
$^3$Institut Mediterrani d'Estudis Avan\c{c}ats, IMEDEA (CSIC--UIB),
E-07071 Palma de Mallorca, Spain \\ 
}

\date{Physical Review E \textbf{59}, 2902--2906, 1999}

\maketitle

\begin{abstract}
We investigate the hierarchical structure of three-frequency resonances in
nonlinear dynamical systems with three interacting frequencies. We hypothesize
an ordering of these resonances based on a generalization of the Farey tree
organization from two frequencies to three. In experiments and numerical
simulations we demonstrate that our hypothesis describes the hierarchies of
three-frequency resonances in representative dynamical systems. We conjecture
that this organization may be universal across a large class of three-frequency
systems.
\end{abstract}


\begin{twocolumns}

Nonlinear systems with two competing frequencies show resonances or lockings,
in which the system locks into a resonant periodic response which has a
rational frequency ratio \cite{huygens}. The locking increases with
nonlinearity, from none in the linear regime, to a critical situation where the
system is everywhere resonant. The subcritical system has quasiperiodic
responses between different lockings, while at supercritical values of the
nonlinearity, chaotic as well as periodic and quasiperiodic responses may
occur. Resonances have been investigated theoretically and experimentally in
many nonlinear systems, and their distribution in parameter space in the form
of a devil's staircase is now well understood, from the number theoretical
concept of Farey trees \cite{oreste,aronson,oreste2,cvit,hao,bogpaper}.
However, all this applies to resonances generated by the interaction of two
frequencies. Far less is known, by comparison, when there are three or more
interacting frequencies. 

Adding another frequency allows new phenomena to take place. Now as well as 
(two-frequency) resonance as before, there is a further possibility:
three-frequency resonance, also known as weak resonance or partial mode
locking. Three-frequency resonances are given by the nontrivial solutions of
the equation $a f_0+b f_1+c f_2=0$, where $a$, $b$, and $c$ are integers, $f_1$
and $f_2$ are the forcing frequencies, and $f_0$ is the resonant response. They
form a web in the parameter space of the frequencies
\cite{baesens,buskirk,cumming,linsay}. In this paper, we hypothesize a local
ordering of three-frequency resonances based on generalizing the Farey tree of
two-frequency systems to three frequencies. We perform experiments and numerical
simulations to show that our hypothesis is justified in representative
dynamical systems with three interacting frequencies: a quasiperiodically
forced circle map, a pair of parametrically coupled forced nonlinear
oscillators, and an experimental system consisting of an electronic circuit of
forced phase-locked loops. Our observations lead us to conjecture that the 
ordering we predict may be universal in a large class of dynamical systems with
three interacting frequencies. 

Firstly, we revise continued fractions and the Farey tree
for the case of two frequencies.
Consider a two-frequency system with autonomous frequency $f_0$ and external 
frequency $f_1$. Let $\tilde f=f_1/f_0$. The aim is to define a sequence of
rational numbers that converges to $\tilde f$. Strong convergence \cite{kinchin} 
is defined for rational fractions $p_i/q_i$, $(p_i, q_i)\in\Z$ as
\begin{equation}
\left\|\tilde f-\frac{p_i}{q_i}\right\|=|q_i\tilde f-p_i|
.\end{equation}
$p_n/q_n$ is a best rational approximation if
\begin{equation}
\left\|\tilde f-\frac{p_n}{q_n}\right\|<\left\|\tilde f-\frac{p_i}{q_i}\right\|
\end{equation}
for all $(p_i,q_i)$ for any $q_i\leq q_n$. Given $\tilde f$, $p_n$ and $q_n$ 
are obtained by expanding $\tilde f$ in continued fractions 
$\tilde f=(a_1,a_2,a_3,\ldots)$, and truncating the expansion as
$p_n/q_n=(a_1,a_2,a_3,\ldots a_n)$ \cite{hardy}.
The $p_n/q_n$ are then the strong convergents of $\tilde f$.
They give the sequence of fractions with lowest monotonically increasing
denominators that converges to $\tilde f$.

The physically motivated hypothesis invoked to explain the local ordering of
the hierarchy of (two-frequency) resonances is that the larger the denominator,
the smaller the plateau. The fraction with smallest denominator between $p/q$
and $r/s$, if they are sufficiently close that $|qr-ps|=1$, when they are
called adjacents, is $(p+r)/(q+s)$. This fraction, known as the mediant, is
then the most important resonance in the interval. Repeatedly performing the
mediant operation
\begin{equation}
\frac{p}{q}\oplus\frac{r}{s}=\frac{p+r}{q+s}
\end{equation}
on a pair of adjacent rational numbers, we obtain a Farey tree. The
Farey tree provides a qualitative local ordering of two-frequency resonances,
and gives rise to a structure of plateaux at all rationals known as the devil's
staircase. The devil's staircase in turn is the skeleton for the layout of the
resonances in parameter space as Arnold tongues 
\cite{oreste,aronson,oreste2,cvit,hao,bogpaper}.

Now consider the case of three frequencies, one internal $f_0$, and two
external $f_1$ and $f_2$. We may divide through by the autonomous frequency 
$f_0$, to give $f_1^\dagger=f_1/f_0$, and $f_2^\dagger=f_2/f_0$. We aim to come
up with two convergent sequences of rational numbers with the same denominator,
$p_n/k_n$ and $q_n/k_n$,  which are strong convergents to $f_1^\dagger$ and
$f_2^\dagger$ respectively. As before, strong convergence is defined as
\begin{equation}
\left\|(f_1^\dagger,f_2^\dagger)-(\frac{p_i}{k_i},\frac{q_i}{k_i})\right\|=
|k_i(f_1^\dagger,f_2^\dagger)-(p_i,q_i)|
.\end{equation}
Thus $(p_n/k_n,q_n/k_n)$ are best rational approximants if
\begin{equation}
\left\|(f_1^\dagger,f_2^\dagger)-(\frac{p_n}{k_n},\frac{q_n}{k_n})\right\|
<\left\|(f_1^\dagger,f_2^\dagger)-(\frac{p_i}{k_i},\frac{q_i}{k_i})\right\|
\end{equation}
for all triplets of integers $(p_i,q_i,k_i)$ for any $k_i\leq k_n$. 
So we may write
\begin{eqnarray}
\varepsilon_1&=&\left\| \frac{p_n}{k_n}-f_1^\dagger\right\|
=|k_n f_1^\dagger-p_n|, \\
\varepsilon_2&=&\left\| \frac{q_n}{k_n}-f_2^\dagger\right\|
=|k_n f_2^\dagger-q_n|
,\end{eqnarray}
where we wish to obtain the integers $p_n$, $q_n$ and $k_n$.
This general problem has not been solved \cite{kim1,kim2}, however, we may set
$\varepsilon_1=\varepsilon_2$, so that both approximations should be equally 
good or bad. If we do this, we can equate
$|k_n f_1^\dagger-p_n|=|k_n f_2^\dagger-q_n|$,
and ask what is $k_n$. There are two solutions
$k_n=(q_n\pm p_n)/(f_2^\dagger\pm f_1^\dagger)$.
At this point we must remember that $k_n$ is an integer, so these solutions
require that the frequencies be rescaled by $f_2^\dagger\pm f_1^\dagger$.
For which we define for the first solution 
$\tilde f_1 = f_1^\dagger/(f_1^\dagger+f_2^\dagger)$, 
$\tilde f_2 = f_2^\dagger/(f_1^\dagger+f_2^\dagger)$,
and similarly for the other solution
$\tilde f_1^* = f_1^\dagger/(f_2^\dagger-f_1^\dagger)$, 
$\tilde f_2^* = f_2^\dagger/(f_2^\dagger-f_1^\dagger)$.
The two solutions give rise to different $\varepsilon$'s
\begin{eqnarray}
\varepsilon  &=&|(p_n+q_n) \tilde f_1-p_n|=|(p_n+q_n) \tilde f_2-q_n|, \\
\varepsilon^*&=&|(q_n-p_n) \tilde f_1^*-p_n|=|(q_n-p_n) \tilde f_2^*-q_n|,
\end{eqnarray}
from which one can obtain
$\varepsilon/\varepsilon^*=|(f_2-f_1)/(f_1+f_2)|<1$.
So in this sense the $(\tilde f_1,\tilde f_2)$ solution is better than the
$(\tilde f_1^*,\tilde f_2^*)$ solution.

Now sticking with the $(\tilde f_1,\tilde f_2)$ solution, $p_n$, $(p_n+q_n)$,
and $q_n$ are obtained from the continued fraction expansions of $\tilde f_1$
and $\tilde f_2$. Since 
\begin{eqnarray}
\tilde f_1 =
\frac{f_1^\dagger}{f_1^\dagger+f_2^\dagger}
=\frac{f_1}{f_1+f_2}=\frac{1}{1+\frac{1}{\frac{f_1}{f_2}}}, \\
\tilde f_2 = \frac{f_2^\dagger}{f_1^\dagger+f_2^\dagger}
=\frac{f_2}{f_1+f_2}=\frac{1}{1+\frac{f_1}{f_2}},
\end{eqnarray}
if we have the continued fraction expansion of $f_1/f_2=(a_1,a_2,a_3,\ldots)$,
that of $\tilde f_1=(a_1+1,a_2,a_3,\ldots)$, and 
$\tilde f_2=(1,a_1,a_2,a_3,\ldots)$
Hence if $p_n/q_n$ is the $n$th strong convergent of $f_1/f_2$, or equivalently of
$\tilde f_1/\tilde f_2$, given by this continued fraction expansion, 
$p_n/(p_n+q_n)$ and $q_n/(p_n+q_n)$ are the strong
convergents of $\tilde f_1$ and $\tilde f_2$ respectively.
If $p_n/q_n$ is a such a convergent of $f_1/f_2$, we may define
as generalized adjacents any pair of $(f_i/r_i$, $f_j/r_j)$, with 
$f\in\R$ and $r\in\Z$, that satisfy
\begin{equation}
|f_i r_j - f_j r_i|=|f_1 q_n - f_2 p_n|
.\end{equation}
From this definition, the subharmonics $f_1/p_n$ and $f_2/q_n$ are
generalized adjacents, and the mediant between them is 
\begin{equation}
\tilde f_s=\frac{f_1+f_2}{p_n+q_n}, 
\end{equation}
which by extension from the two-frequency case we hypothesize to be the largest
plateau between $f_1/p_n$ and $f_2/q_n$. Starting
with $(\tilde f_1^*,\tilde f_2^*)$ instead of $(\tilde f_1,\tilde f_2)$, we
obtain the generalized mediant $\tilde f_s^*=(f_2-f_1)/(q_n-p_n)$; both $\tilde
f_s$ and $\tilde f_s^*$ are shown in Fig.~\ref{farey}(a). The generalized
mediant operation
\begin{equation}
\frac{f_1}{r_i}\oplus\frac{f_2}{r_j}=\frac{f_1+f_2}{r_i+r_j}
\label{mediant}\end{equation}
then provides us with a generalized Farey tree for three
frequencies; in Fig.~\ref{farey}(b) we show the first three levels of the
tree obtained by recursive application of Eq.~(\ref{mediant}) to the
adjacents $f_1/p_n$ and $\tilde f_s^*$. 

\begin{figure}[tb]
\begin{center}
\def\epsfsize#1#2{0.46\textwidth}
\leavevmode
\epsffile{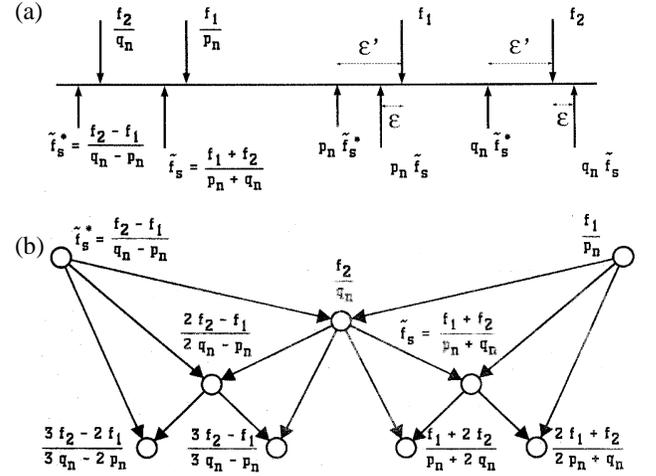}
\end{center}
\caption{
(a) Schematic diagram of the frequency line, showing 
the relative position of $\tilde f_s $ and $\tilde f_s^*$ and their respective
$p_n$, $q_n$ harmonics that approximate $f_1$ and $f_2$ at equal distance. 
(b) Generalized Farey tree starting from the
adjacents $f_1/p_n$ and $\tilde f_s^*$. The first mediant is $f_2/q_n$; at the
second level we obtain $\tilde f_s$ and $(2f_2-f_1)/(2q_n-p_n)$, and so on.}
\label{farey}\end{figure}

Let us take as an example a three-frequency system with the two external 
frequencies set to $f_1=2100$~Hz and $f_2=3600$~Hz. The frequency ratio
$f_1/f_2$ is then 7/12. The continued fraction expansion for $f_1/f_2$ is
$(1,1,2,1,1)$, and the different truncations of this produce the convergents of
7/12, which are 1/1, 1/2, 3/5, and 4/7. So we take 4/7 as an approximation to 
the higher-order rational 7/12, or equivalently in terms of the original
frequencies, $f_1/f_2=$2000~Hz/3500~Hz approximates $f_1/f_2=$2100~Hz/3600~Hz
with a detuning of 100~Hz. Between the adjacents $f_1/4=525$~Hz and
$f_2/7=514.3$~Hz lies the mediant $\tilde f_s=(f_1+f_2)/11=518.2$~Hz, which we
have hypothesized to be the widest resonance in this interval, and recursively
applying the mediant operation gives us the Farey tree that predicts the entire
hierachy of resonances in the interval. We now proceed to test this hypothesis
in three-frequency dynamical systems.

\begin{figure}[tb]
\begin{center}
\def\epsfsize#1#2{0.46\textwidth}
\leavevmode
\epsffile{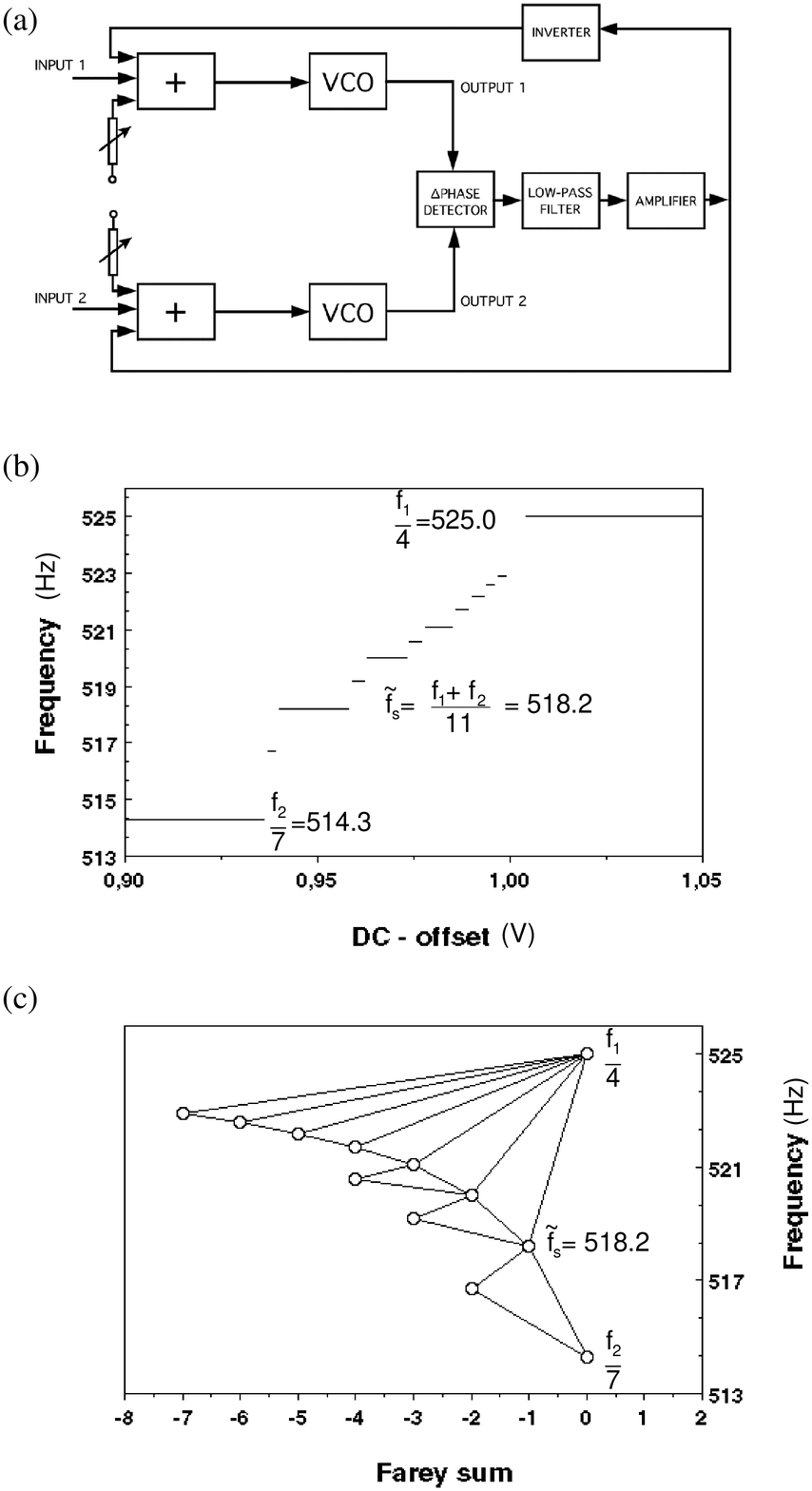}
\end{center}
\caption{
(a) The electronic circuit in block-diagram form. Two coupled 
voltage-controlled oscillators (VCOs) are forced with two independent external 
forces.
(b) Three-frequency devil's staircase for the circuit. The straight lines
correspond to intervals of the DC offset within which the fundamental frequency
of the output remains constant. We varied the DC offset of input 1 in steps of
1 mV in the interval 0.90~V to 1.05~V, which corresponds to frequency responses
of the circuit between 514.28 Hz and 525.00 Hz. Only intervals with a stability
width greater than 2 mV are plotted. The external frequencies are fixed at
2100 Hz and 3600 Hz.
(c) The generalized Farey tree predicts the organization of all the  
frequency values observed in (b).}
\label{circuit}\end{figure}

We have constructed the three-frequency electronic oscillator shown in
Fig.~\ref{circuit}(a) \cite{ourosc}. Two phase-locked loop (PLL) circuits made
up of voltage-controlled oscillators (VCOs) are coupled through a lowpass and
integrator network and forced with two independent periodic forces. Typically,
the outputs of the two oscillators are synchronized 1/1. We filter the outputs
of the two oscillators in order to attenuate the components at the external
frequencies and we measure the output frequencies of both oscillators. If the
two values coincide we plot them against the DC offset of oscillator 1, which
is used as the control parameter (for the circuit a variation in the mean value
of the $i$th external force is equivalent to a linear change in the natural
frequency of the $i$th oscillator). The results are presented in
Fig.~\ref{circuit}(b) for the interval $(f_2/7,f_1/4)$ for $f_1$ and $f_2$
fixed at 2100 Hz and 3600 Hz respectively. This appears to be a typical devil's
staircase familiar from periodically driven oscillators. It is, however, a
three-frequency devil's staircase: the plateaux correspond to solutions with
three linearly dependent basic frequencies (the frequency plotted plus the two
forcing frequencies) instead of to periodic solutions. As predicted, the
generalized mediant $\tilde f_s=518.2$~Hz is the largest resonance in the
interval, moreover, the generalized Farey tree shown in Fig.~\ref{circuit}(c) 
gives the entire hierarchy of resonances of Fig.~\ref{circuit}(b).

\begin{figure}[tb]
\begin{center}
\def\epsfsize#1#2{0.46\textwidth}
\leavevmode
\epsffile{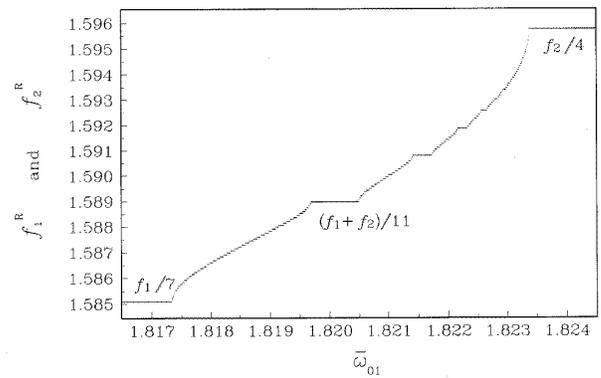}
\end{center}
\caption{
The three-frequency devil's staircase for the system of ordinary differential
equations of Eq.~(\ref{pirogon}). The system parameters are equivalent to those
of Fig.~\ref{circuit}(b).}
\label{pg_out}\end{figure}

We have integrated an exactly soluble system of ordinary differential equations
with three interacting frequencies \cite{oreste,oreste2}
\begin{eqnarray}
\ddot x_i+(4b x_i^2-2a)\dot x_i+b^2 x_i^5-2ab x_i^3& \nonumber \\
\mbox{}+(\omega_{0i}(t)^2+a^2)x_i&=f_i(t)
\label{pirogon}\end{eqnarray}
for $i=1,2$. The external forces and the coupling term are chosen in such a way
as to preserve the piecewise integrability of the overall system. The
oscillators are coupled parametrically, their intrinsic frequency changing
every time the coordinate of one of them changes sign 
$\omega_{0i}(t)=
\tilde\omega_{0i}+{\mathrm sgn}\,u_{i}(t)\,{\mathrm sgn}\,u_{j}(t)\Delta_i$. 
The $i$th oscillator is driven by an impulsive external
force $f_i(t)=V_{Ei} \sum_n\delta(t-n\tau_{Ei})$ of frequency $\omega_i$ whose
effect is to produce a discontinuity $V_{Ei}$ in the oscillator velocity at
times $n\tau_{Ei}=n/(2\pi\omega_i)$. We have performed a power spectrum
analysis of the output of both oscillators while varying the intrinsic
frequency of oscillator 1. We display in Fig.~\ref{pg_out} the most
prominent peak in each spectrum against $\tilde\omega_{01}$ for a
parameter region equivalent to that of Fig.~\ref{circuit}(b), and in which the
two oscillators are also synchronized 1/1. All detectable resonances are again
well described by the Farey tree structure of Fig.~\ref{circuit}(c). 

Our final example of a three-frequency system is the quasiperiodically forced
circle map
\begin{equation}
\phi'=\phi+\Omega_n+{\frac{k}{2\pi}}\sin 2\pi\phi\bmod 1
.\end{equation}
The quasiperiodic sequence $\Omega_n$ is the time interval between successive
pulses of a sequence composed of the superposition of two periodic
subsequences, one of period equal to one and the other of period $T_1/T_2 < 1$
(with no loss of generality), multiplied by the value of the intrinsic frequency
of the oscillator. In Fig.~\ref{circle}(a) we demonstrate that, for the same
input frequencies as in the previous two cases, the output in the form of a
devil's staircase is qualitatively unchanged; once again its organization is
given by the generalized Farey tree. Moreover, Fig.~\ref{circle}(b) shows how
three-frequency resonances are arranged globally in the form of devil's ramps
in the parameter space.

By generalizing the known Farey tree structure of two-frequency resonances for
three frequencies we have predicted the organization of three-frequency
resonances in dynamical systems with three interacting frequencies. Our results
for three different three-frequency dynamical systems for the same ratio of
forcing frequencies $f_1/f_2$ show that in each example the predicted
generalized Farey tree hierarchy is observed. We have repeated these
experiments and simulations for different frequency ratios, both rational, as
the example presented here, and irrational, for example
$f_1/f_2=(1+\sqrt{5})/2$, the golden ratio. In every case we have examined, the
local ordering of resonances around the convergents in the devil's staircase
--- slices through the devil's ramps of Fig.~\ref{circle}(b) --- is well
described by the generalized Farey tree hierarchy. We conclude that the
organization we have described here is widespread, and conjecture that it may
be universal in a large class of three-frequency systems.

JHEC and OP acknowledge the financial support of the Spanish Direcci\'on 
General de Investigaci\'on Cient\'\i fica y T\'ecnica, contracts PB94-1167 and 
PB94-1172.

\begin{figure}[tb]
\begin{center}
\def\epsfsize#1#2{0.46\textwidth}
\leavevmode
\epsffile{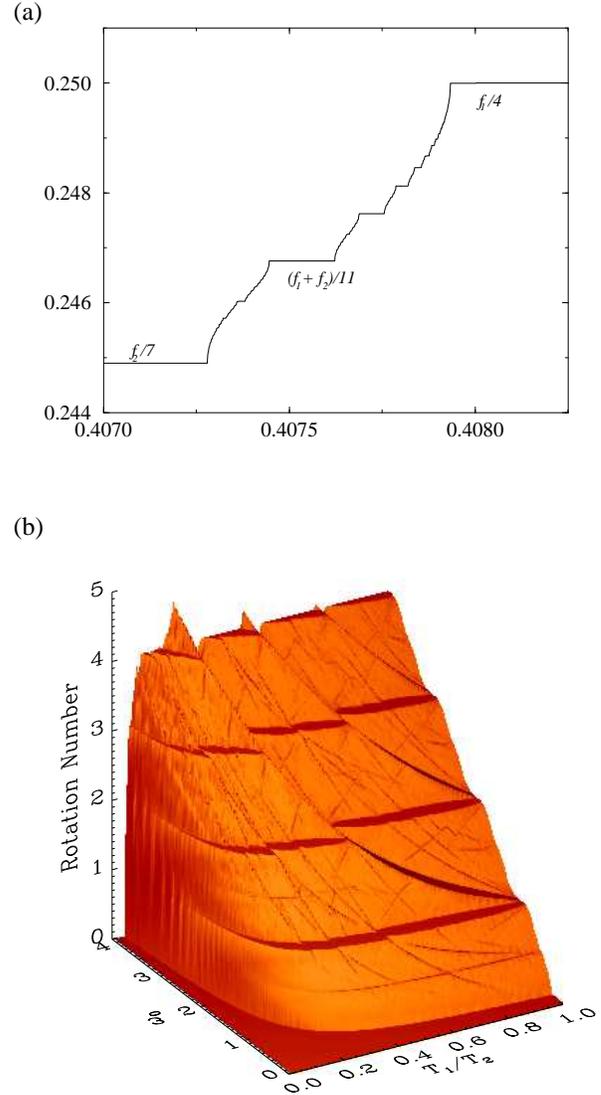}
\end{center}
\caption{
Quasiperiodically forced circle map:
(a) Three-frequency devil's staircase for forcing frequencies $f_1=1$,
$f_2=12/7$. 
(b) Devil's ramps: the rotation number as a function of the external
frequency ratio and the intrinsic frequency shows the global organization of 
three-frequency resonances.}
\label{circle}\end{figure}

\end{twocolumns}
\end{document}